# The environment of HII galaxies

Eduardo Telles[1,2]
Roberto Terlevich[2]
1. *Institute of Astronomy, Madingley Road, Cambridge CB3 0HA, U.K.*
2. *Royal Greenwich Observatory, Madingley Road, Cambridge CB3 0EZ, U.K.*
*etelles@mail.ast.cam.ac.uk & rjt@mail.ast.cam.ac.uk*

11 January 1995

**ABSTRACT**

Recent morphological studies (Telles & Terlevich 1994) of HII galaxies, i.e. dwarf galaxies dominated by a very luminous starburst, have indicated that luminous HII galaxies tend to show distorted morphology suggestive of tidal interactions triggering the present starburst while low luminosity HII galaxies tend to be instead symmetric and regular. To check the tidal origin of the starburst in HII galaxies, we have searched for companions in the neighbourhood of a sample of 51 HII galaxies. We found that only 12 HII galaxies have a neighbour within a projected distance of 1 Mpc and 250 $\mathrm{km\,s^{-1}}$ in velocity difference, and of these 12, only 4 have a luminous ($M_B < -19$) neighbour. Surprisingly, isolated HII galaxies tend to be of high luminosity and disturbed morphology while HII galaxies with neighbours tend to be low luminosity regular HII galaxies. Furthermore, the metal abundance and the equivalent width of the emission lines in HII galaxies do not depend on the presence of a companion.

These results are opposed to simple expectations if interaction with a bright companion is the main mechanism triggering the starbursts.

We have also found a loose group of HII galaxies with no luminous companion. For this, there is the additional difficulty of understanding how these starbursts are synchronized on time scales of less than $10^7$ yrs in systems separated by $\sim$ 1-2 Mpc.

## 1 INTRODUCTION

It is generally accepted that bursts of star formation in galaxies may be triggered by interaction with other galaxies or that the star formation activity in galaxies may be substantially enhanced by galaxy interactions (Sanders et al. 1988). Clear cases in which interaction among galaxies may have produced starbursts, cannibalism or tidal disruption of the systems tend to show distorted light distribution and peculiar morphologies, e.g. the extreme case of ultraluminous IRAS galaxies (Melnick & Mirabel 1990). Isolation, compactness or absence of morphological peculiarities in starburst galaxies are usually taken as evidence *against* interactions being the triggering mechanism.

Among starburst galaxies, HII galaxies are probably the youngest systems that can be studied in any detail. They are galaxies dominated by a starburst and most of them were found in objective prism surveys because of their strong emission lines (Terlevich 1988). Their low metallicity, present high star formation rate (SFR) and total content of neutral hydrogen put constraints on the past history of star formation and the number of bursts these systems may have undergone (Thuan 1991). Although the general spectroscopic properties are fairly uniform among HII galaxies, they seem to differ significantly in their morphological and environmental properties (Melnick 1987, Telles 1994). Our high resolution imaging study of a sample of HII galaxies (Telles & Terlevich 1994) has revealed that the luminous ($M_V < -18$)[*] HII galaxies, hereafter called type I HII galaxies, show almost invariably evidence of distorted morphologies, tails, plumes or strong asymmetries in their CCD images while the low luminosity ($M_V > -18$) HII galaxies, hereafter called type II HII galaxies, tend to be single, regular, symmetric and compact. This result strongly suggests a close relation between luminosity and distorted morphology.

Recent work has raised the possibility that collisions between giant galaxies eject matter into the intergalactic medium from which dwarf galaxies may be formed (Mirabel 1992; Mirabel et al. 1992, 1994; Duc & Mirabel 1994). In this scenario, star forming regions generated on the tips of tidal tails that emanate from merging disk galaxies may become gravitationally bound systems that detach from the merging system and resemble HII galaxies.

---

[*] We have used $H_0 = 50$ $\mathrm{km\,s^{-1}\,Mpc^{-1}}$ throughout this work.

In order to investigate the possible relation between the distorted morphology of type I HII galaxies and tidal interactions, and put constraints on the number of HII galaxies that may be formed this way, we have analysed the environment of a sample of 51 HII galaxies for which we have high resolution CCD images, redshifts, luminosities, emission line ratios, emission line widths and accurate chemical abundances. Our working hypothesis is that if the violent star formation activity of the HII galaxies has been triggered by an encounter or an interaction with a massive galaxy we expect to find a luminous galaxy in the proximity of the HII galaxies.

Early investigations have attempted to identify sub-samples of HII galaxies presenting more homogeneous morphological and environmental properties (Loose & Thuan 1985, Melnick 1987, Kunth et al. 1988, Salzer et al. 1989). Melnick (1987), for instance, reported that 50 % of a sample of 50 HII galaxies seem to be isolated. His conclusion was based solely on the presence of an obvious interacting companion inside the CCD field of each object. Campos-Aguilar & Moles (1991) have searched for bright companions using the CFA catalogue within a volume defined by 1 Mpc of projected distance and a difference in radial velocity of 500 km s$^{-1}$ around each HII galaxy. They concluded that 1/3 of the 48 most compact HII galaxies from their total sample of 71 objects seem to be isolated. In a similar work Campos-Aguilar et al. (1993) confirmed their earlier results by finding that 64% of a sample of 85 galaxies from the University of Michigan lists IV and V are devoid of any neighbour and only 11 galaxies show a massive companion within that volume but none close enough to be visibly associated with the star forming dwarf galaxy in question. Salzer & Rosenberg (1994), Rosenberg et al. (1994) and Pustil'nik et al. (1994) have analysed the spatial distribution of several samples of HII galaxies. They have all reached identical conclusions that HII galaxies are less clustered than "normal" galaxies, most are isolated and a few are found in voids (i.e. the distance to the nearest neighbour is greater than 5 h$^{-1}$ Mpc).

In this paper we report a search for *giant* companions from the latest compiled database of galaxies available in the NASA Extragalactic Database. The compilation is essentially complete to B = 14.5 mag. In order to further investigate the relation between tidal interactions and triggering of star formation, we searched for correlations between local environment and the intrinsic properties of HII galaxies such as luminosity, metallicity, morphology, equivalent width of H($\beta$) and velocity dispersion. In §2 we describe our data sample and methods. We present our results in §3. Finally, we discuss our results and conclusions in §4.

## 2  THE DATA AND METHODS

### 2.1  The Sample

The sample used in the present study has been used by Melnick, Terlevich & Moles (1988) in the application of HII galaxies as distance indicators and more recently in our re-analysis of the morphology and dynamics of HII galaxies (Telles & Terlevich 1993, Telles 1994). The sample is not representative of the whole data base of dwarf emission line galaxies used in various other studies as it includes only bright HII galaxies with large values of W(H$\beta$) ($>$ 30 Å). The brightness criterion was adopted in order to facilitate their determination of the emission line profiles, while objects with large W(H$\beta$) were selected to avoid both evolved bursts and bursts with contamination from an older population. The sample constitutes a fairly homogeneous sub-sample of *the spectrophotometric catalogue of HII galaxies* (Terlevich et al. 1991, hereafter SCHG). The spectroscopic information for this sub-sample can be found in Melnick et al. (1988) and in SCHG. Information on the morphology of all objects can be found in Telles (1994) who analysed CCD direct images from the 1.54m Danish telescope in La Silla, Chile.

### 2.2  The selection criteria and method

We have used the information compiled in the NASA/IPAC Extragalactic Database (NED) [†] in search of bright companions. This data base contains a large quantity of data in different wave bands from various astronomical catalogues presently available in the literature. The completeness of the NED depends on the completeness of the individual catalogues which are included in the database. For equatorial or northern objects ($\delta >$ -3°30 min) the Zwicky catalogue (CGCG, Zwicky 1961) is complete to B$_{lim} \approx$ 15.5 and for southern objects ($\delta <$ -17°30 min) which comprises most of our present sample, the relevant catalogues are the ESO Quick (B) (Lauberts 1982) and its more recent automated version the ESO-LV catalogue (Lauberts & Valentijn 1989). Despite the fact that the ESO catalogues are diameter limited, they are effectively complete down to B$_{lim} \leq$ 14.5 for all types (Lauberts & Valentijn 1989, Hudson 1992). We have, therefore, improved the work of Campos-Aguilar & Moles (1991), who used the CFA catalogue with a magnitude limit of B $\leq$ 13.5, by lowering the magnitude limit of the search by 1.0 magnitude conservatively considering galaxies of all types, and possibly 1.5 magnitudes in the case of late type galaxies, preferentially selected in diameter limited catalogues (Lauberts & Valentijn 1989, Hudson 1992).

The criteria we used to establish the search radius follows from the need to test the hypothesis that the starbursts in HII galaxies are triggered by tidal interactions. Following Icke (1985), the pericentre distance ($d$) below which shocks can be expected by a tidal interaction with a distant perturber leading to star formation (Zaritsky & Lorrimer 1993) is given by:

$$d \sim (8\pi\mu \ (v/s_0))^{\frac{1}{3}} \times R$$

where $\mu$ is the ratio of the mass of the intruder over the mass of the victim galaxy; $v$ is the circular speed velocity at the galactic radius $R$ and $s_0$ is the sound speed of the ISM. Assuming rough numbers for these parameters we find $d$ to be of the order of a few hundred Kpc, but most certainly less than 1 Mpc. This value represents the maximum distance one may simple mindedly expect to find a perturber galaxy capable of triggering star formation. We have therefore adopted as companion criterion in our searches any galaxy inside a projected distance of 1 Mpc of a given HII galaxy and with a difference in radial velocity smaller than 250 km s$^{-1}$.

---

[†] The NASA/IPAC Extragalactic Database (NED) is operated by the Jet Propulsion Laboratory, California Institute of Technology, under contract with the National Aeronautics and Space Administration

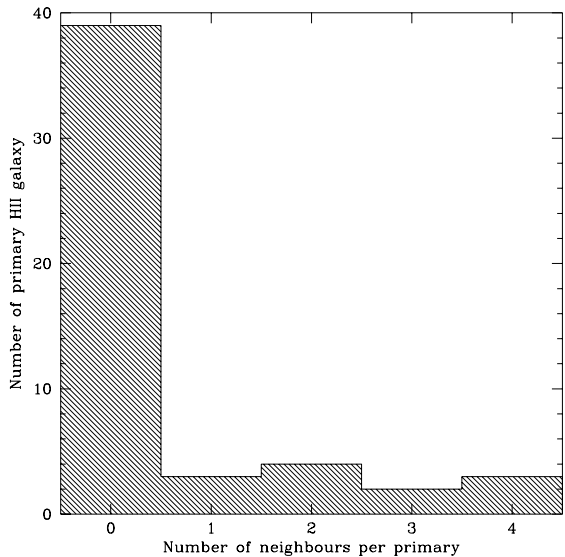

**Figure 1.** Number of bright companions within 1 Mpc of projected distance and 250 km s$^{-1}$ velocity difference from an HII galaxy

## 3 RESULTS AND DISCUSSION

Figure 1 shows the histogram of all 51 objects and the number of all bright companions that have been found in the specified volume for each of them. We find that 39 HII galaxies out of 51 have no bright companion.

Table 1 presents the data for the 12 HII galaxies found to have at least one other galaxy in its neighbourhood (within 1 Mpc of projected distance and 250 km s$^{-1}$ in radial velocity difference) and the information about their neighbours. We have assumed that the observed redshift represents the cosmological radial velocity as distance in Euclidean space. Columns 1-6 give the name of the HII galaxy and its spectroscopic information. Columns 7-12 give the name, type, apparent and absolute blue magnitudes, projected angular distances (in arcminutes) and finally the spatial physical distances (in Kpc) of bright objects in the vicinity of the primary HII galaxy (a negative distance means that the neighbour is a foreground galaxy).

Important results can be drawn directly from table 1:
1) We find a loose group of HII galaxies (UM 461, UM 462, UM 463, UM 465) which are neighbours among themselves within the specified cylindric volume and have *no* associated bright neighbour. 2) Only 4 objects (To1008-286, Fairall 30, Mrk 36, To1334-326) have one relatively luminous (M$_B$ < -19) galaxy within 1 Mpc$^3$. 3) No companion is brighter than $M_B^* \sim -21.5$. 4) Even restricting the analysis to the neighbours brighter than the magnitude limit for which our search is complete for all types of galaxies, we note that virtually all neighbours are late Hubble type galaxies.

None of the above results, however, imply any physical relationship between these few HII galaxies and their neighbours. In fact, a close analysis of the PSS prints found no evidence of interaction in the companions listed in table 1 or any visible association with the respective HII galaxy. These results do not rule out the hypothesis that some dwarf galaxy sized systems may be formed and detach from a strongly interacting pair of giant galaxies as suggested by Mirabel et al. (1992), but indicates that the majority of HII galaxies are probably *not* formed that way.

We have also investigated if the properties of HII galaxies with listed bright companions in table 1 differ from the majority of HII galaxies without bright neighbours. Figure 2 presents a set of histograms showing the distribution of their spectroscopic properties: H$\beta$ luminosity (L(H$\beta$) corrected for extinction using the Balmer decrement as the reddening indicator), emission line width or velocity dispersion ($\sigma$), H$\beta$ equivalent width (W(H$\beta$)) and oxygen abundance (12+log(O/H)) of HII galaxies. The histograms are divided in two sets, the objects with bright companions are represented by the histograms with hatched lines while those in thick lines represent the subset of isolated HII galaxies.

There is a clear trend for the HII galaxies with bright companions to have lower L(H$\beta$) and smaller $\sigma$, in agreement with the known correlation between L(H$\beta$) and $\sigma$ (Terlevich & Melnick 1981; Melnick, Terlevich & Moles 1988) observed in HII galaxies. On the other hand the distributions of W(H$\beta$) and oxygen abundance in figure 2 seem to be quite similar for galaxies with and without bright companions. This implies that if a giant companion has taken part in any respect (if at all) in the onset of the present burst of star formation by acting as distant external perturber of the ISM of the primary HII galaxy, it does not determine the starburst chemical and stellar evolution. In addition, none of the spectroscopic properties (L(H$\beta$), $\sigma$, W(H$\beta$), O/H) of the HII galaxies correlates with the properties of the nearest bright object (distance, absolute magnitude) reinforcing the view that the bright neighbours of HII galaxies in groups are not related with the possible causes for the triggering of their present burst of star formation.

We have checked the trend shown in figure 2 by performing a Kolmogorov-Smirnov test to our unbinned distributions in order to assess statistically whether the spectroscopic data for HII galaxies with neighbours and isolated are similar or not, as discussed above. The low values of the significance level ($P \approx 0.01$) disproves that the distributions of L(H$\beta$) and $\sigma$ can be drawn from the same distribution function. However, the high values of the significance level ($P > 0.5$) for the distributions of W(H$\beta$) and O/H indicate that the data are consistent with them being drawn from a single distribution function.

As mentioned before, our morphological study of the present sample of HII galaxies (Telles & Terlevich 1994, Telles 1994) indicated that all high luminosity HII galaxies show signs of distorted morphologies in the form of tails, wisps or fans while low luminosity HII galaxies tend to be compact and regular in shape. We therefore reach an apparent contradiction: HII galaxies with companions are of

**Table 1.** Properties of the HII galaxies and their bright neighbours

| HII galaxy | Z | $\sigma$ | $\mathcal{F}$ (H$\beta$) | W (H$\beta$) | O/H | Bright neighbour | Type | B mag | $M_B$ Mag | proj-dist arcmin | tot-dist Kpc |
|---|---|---|---|---|---|---|---|---|---|---|---|
| To1004-294 | 0.004 | 30.6 | 2.70 | 60 | 8.23 | ESO435-G016 | I0pec | 13.41 | -17.91 | 140 | 789 |
|  |  |  |  |  |  | TOLOLO0957-278 | EmLS | 14.38 | -16.38 | 152 | -3839 |
|  |  |  |  |  |  | NGC3113 | SAB(s)d | 13.33 | -18.36 | 95 | 3878 |
|  |  |  |  |  |  | NGC3137 | SA(s)cd | 12.10 | -19.63 | 64 | 4215 |
| To1008-286 | 0.014 | 24.0 | 0.30 | 125 | 8.16 | MCG-05-24-025 | E? | 14.43 | -20.20 | 8.2 | 475 |
| Fairall30 | 0.004 | 21.8 | 1.80 | 90 | 8.01 | CGCG1049+0451 | Scd: | 14.00 | -17.59 | 117 | -731 |
|  |  |  |  |  |  | NGC3423 | SA(s)cd | 11.59 | -19.94 | 77 | -887 |
| MK36 | 0.002 | 16.0 | 1.40 | 70 | 7.86 | NGC3486 | SAB(r)c | 11.05 | -19.62 | 70 | 757 |
|  |  |  |  |  |  | CGCG1059+2858 | Im | 14.90 | -15.83 | 55 | 1077 |
|  |  |  |  |  |  | NGC3413 | S0 | 13.08 | -17.47 | 299 | -1050 |
|  |  |  |  |  |  | NGC3510 | SB(s)m | 12.70 | -18.05 | 24 | 1183 |
| To1116-325 | 0.002 | 12.0 | 0.43 | 275 | 8.31 | NGC3621 | SA(s)d | 10.18 | -20.63 | 5.2 | 2560 |
|  |  |  |  |  |  | ESO439-G012 | Sa | 16.64 | -14.24 | 254 | 3143 |
| To1147-283 | 0.006 | 18.9 | 0.24 | 45 | 7.90 | DDO239 | SB(s)d: | 12.97 | -19.97 | 30 | -3277 |
| To1324-276 | 0.006 | 33.4 | 0.87 | 115 | 8.20 | ESO509-G026 | SAB(s)m | 15.17 | -17.66 | 30 | 553 |
|  |  |  |  |  |  | NGC5101 | SB(rs)0/a | 11.60 | -21.25 | 86 | 1254 |
| To1334-326 | 0.013 | 16.4 | 0.3 | 265 | 8.00 | ESO383-G037 | SB0? | 14.04 | -20.22 | 17 | -742 |
|  |  |  |  |  |  | ESO383-G048 | S0 | 14.53 | -19.87 | 38 | 4296 |
|  |  |  |  |  |  | NGC5215A | S0pec | 14.10 | -20.32 | 45 | 4905 |
| To1457-262 | 0.018 | 51.5 | 1.0 | 95 | 8.19 | ESO-LV5130110 | G | 15.85 | -19.17 | 0.6 | -1120 |
| UM461 | 0.003 | 14.5 | 0.70 | 155 | 7.74 | UM462 | EmLS | 14.58 | -16.95 | 17 | 2262 |
|  |  |  |  |  |  | UM465 | EmLS | 14.14 | -17.54 | 155 | 3865 |
| UM462 | 0.003 | 18.5 | 0.75 | 75 | 7.98 | UM461 | EmLS | 16.35 | -14.92 | 17 | -2262 |
|  |  |  |  |  |  | UM463 | EmLS | 17.87 | -14.19 | 108 | 3804 |
|  |  |  |  |  |  | UM465 | EmLS | 14.14 | -17.54 | 158 | 1784 |
| UM465 | 0.003 |  |  | 0.28 | 10 | UM462 | EmLS | 14.58 | -16.95 | 158 | -1784 |
|  |  |  |  |  |  | UM463 | EmLS | 17.87 | -14.19 | 52 | 2267 |
|  |  |  |  |  |  | CGCG1152.6+0200 | G | 14.90 | -17.18 | 96 | 4470 |
|  |  |  |  |  |  | UM461 | EmLS | 16.35 | -14.92 | 155 | -3865 |

Spectroscopic data are from Melnick et al. (1988): redshift (Z), the gas velocity dispersion ($\sigma$ in km s$^{-1}$), the H$\beta$ flux ($\mathcal{F}$(H$\beta$)) in units of $10^{-13}$ erg cm$^2$ s$^{-1}$), the H$\beta$ equivalent width (W(H$\beta$) in Å), and the oxygen abundance (O/H in units of 12+log(O/H)). We adopt $H_0 = 50$ km s$^{-1}$ Mpc$^{-1}$.

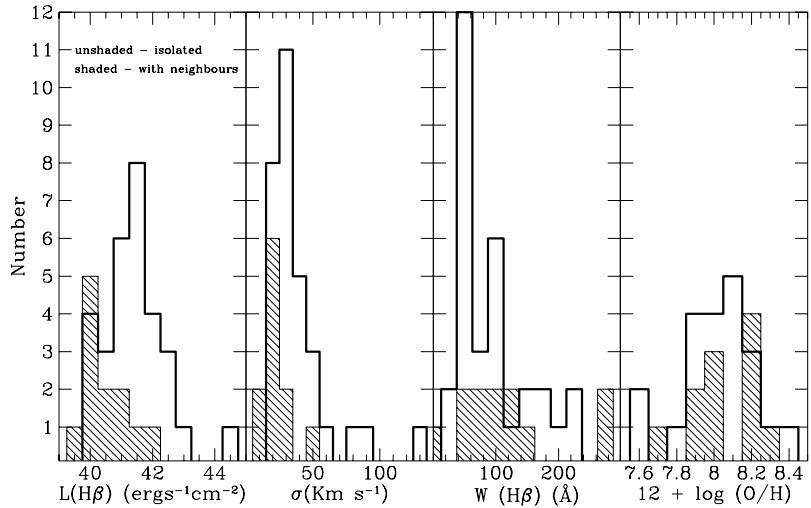

**Figure 2.** Distributions of H$\beta$ luminosity L(H$\beta$), velocity dispersion ($\sigma$), H$\beta$ equivalent width W(H$\beta$) and oxygen abundance 12+log(O/H) for HII galaxies with companions (hatched histograms) and isolated (thick line histograms).

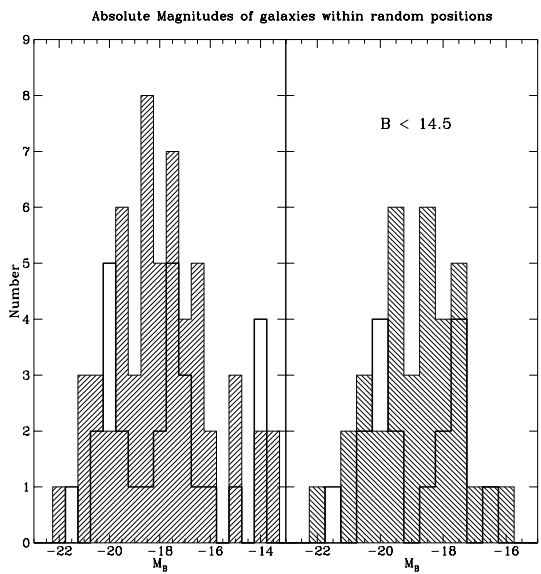

**Figure 3.** Left panel: All bright companions found for a volume in space of 1 Mpc of projected distance and 250 km s$^{-1}$ velocity difference from a random position (shaded histogram) and for our sample of HII galaxies (solid line histogram). Right panel: the same as left panel but for B < 14.5. $H_0 = 50$ km s$^{-1}$ Mpc$^{-1}$.

low luminosity and show little evidence of distortions while isolated HII galaxies are brighter and have distorted morphology. These facts are difficult to understand in a scenario where the starbursts are triggered solely by tidal interaction with a bright companion.

### 3.1 Comparison with random positions and with Sc galaxies

The purpose of this exercise is to check some of the clustering properties of HII galaxies, that is, whether the neighbours of HII galaxies found are chance occurrences or if indeed some HII galaxies tend to be in groups or clusters. We have searched for galaxies within a volume in space comprised of a 1 Mpc radius cylinder and 250 km s$^{-1}$ in radial velocity difference from a random position, keeping the same redshift distribution and search volume as for the search for neighbours of HII galaxies. We then randomly chose 500 positions with some restrictions, namely the range of declination should be about the same as that of the HII galaxies to avoid the choice of positions where there are no reliable catalogues ($-17^\circ < \delta < -3^\circ$) and all positions must fall at least 15 degrees from the galactic plane.

Figure 3 shows the histogram of absolute magnitude for all galaxies within the search volume (1 Mpc radius and 250 km s$^{-1}$ in radial velocity difference). The hatched histogram is the search in random positions, while the thick line histogram represents all neighbours of HII galaxies within this volume unit. The histograms on the right panel are the same but restricting the analysis to galaxies brighter than the magnitude limit for completeness. These histograms show that the companions found for some HII galaxies and the random sample representing the field galaxies have similar luminosity distribution. This suggests that the luminosity of the companion is not affected by the presence of the HII galaxy.

In analogy with the exercise of searching for galaxies in random positions in the sky, we have also performed a Monte-Carlo test using a sample of normal late type galaxies as targets in search for bright neighbours within the same volume and redshift distribution as for the search for neighbours around HII galaxies. We have chosen a sample of 60 Sc galaxies randomly selected from the ESO-Uppsala and RC3 catalogues of galaxies. Although we are unable to estimate the probability of finding companions from the 2-point correlation function because we do not really know the selection function of the galaxies found in NED, we can still ask if HII galaxies have more or fewer companions than Sc galaxies.

Table 2 compares the results of our search for neighbours of HII galaxies with the results of our search for galaxies within the same volume from a random position as well as with the results of our search for neighbours of normal Sc galaxies. For HII galaxies, 12 have been found to have in total 29 neighbours as shown in table 1; 10 HII galaxies have neighbours (17 in total) brighter than B≤14.5. From what follows we additionally exclude the HII galaxy group (UM461, UM462, UM465), thus our final result is that 8 HII galaxies have 15 neighbours brighter than B≤14.5.

Column 5 in table 2 gives the results for the random search irrestrictive of apparent magnitudes and column 6 gives the results for galaxies brighter than B≤14.5. We find 21 volume units out of 500 random positions with at least one galaxy (B ≤ 14.5), that is only ∼4 %. The total number of galaxies found in these 21 cylinders (1 Mpc radius and 250 km s$^{-1}$ in velocity difference) is 36. For our control sample of Sc galaxies we find 17 objects with at least one bright neighbour (B ≤ 14.5) as shown in the last column of table 2 (∼28 %). The total number of neighbours for these 17 Sc galaxies (within the same volume as before) is 43. These may be compared with the 8 HII galaxies which have at least one neighbour (B ≤ 14.5, and excluding a possible local bias of the UM 462 group) out of 51 searches. This is about 16 %. The total number of neighbours of these 8 HII galaxies is 15.

Therefore the total number of galaxies found per Mpc$^3$ is

$$\phi = \frac{N}{n \times \pi \times r^2 \times l}$$

where $N$ is the total number of galaxies found, $n$ is the number of volume units (positions) and $r$ and $l$ are the cylinder radius and length. For our random sample $\phi = 2 \times 10^{-3}$ gals/Mpc$^3$. For our control sample of Sc galaxies $\phi = 2 \times 10^{-2}$ gals/Mpc$^3$. For our sample of HII galaxies we find $\phi = 9 \times 10^{-3}$ gals/Mpc$^3$. Although this result suffers from small number statistics, the excess of number density of galaxies around HII galaxies in comparison with the random search shows that HII galaxies are clustered. In addition, the fact that the number density of galaxies around HII galaxies is a factor of two smaller than what is found for Sc galaxies is an indication that HII galaxies tend to populate low density environments in agreement with other works (Salzer 1989, Vilchez 1994, Rosenberg & Salzer 1994).

Table 2. Results of the search for neighbours and galaxies within the same volume from random positions and from normal Sc galaxies.

| total number of searches | HII Galaxies 51 | | | Random 500 | | Sc Galaxies 60 | |
|---|---|---|---|---|---|---|---|
| | B ≤ 14.5 | | | B ≤ 14.5 | | B ≤ 14.5 | |
| # of searches with neighbours | 12 | 10 | 8 | 30 | 21 | 27 | 17 |
| total # of neighbours found | 29 | 17 | 15 | 57 | 36 | 73 | 43 |

## 4 CONCLUSIONS

Following our findings that bright HII galaxies tend to have evidence of multiple or distorted morphology while low luminosity ones are single and regular, we have analysed the relation between morphology and environment for a sample of 51 HII galaxies for which we have CCD images, redshifts, luminosities, emission line ratios, emission line widths and accurate chemical abundances. Our search for companions for these HII galaxies shows that:

(i) Excluding the UM462 group, less than 20% of the HII galaxies in our sample have a neighbour within a cylindric volume of 1 Mpc in projected distance and 250 km s$^{-1}$ in velocity difference.

(ii) No neighbour found is more luminous than $M_B^* \sim$ −21.5.

(iii) Only less than 10% of HII galaxies have relatively luminous ($M_B <$−19) and close enough ($<$ 1 Mpc) companions which could have acted as a distant perturber to trigger their present burst of star formation.

(iv) HII galaxies with neighbours tend to be those of lower H$\beta$ luminosity and lower velocity dispersion.

(v) The distribution of metallicity (O/H) and H$\beta$ equivalent width of (W(H$\beta$)) are the same for isolated and HII galaxies with neighbours.

(vi) The number density of galaxies is $\approx$ 4 times higher around HII galaxies than would be expected if they were a uniformly distributed sample but a factor of two lower than the environment of normal late-type spiral galaxies.

(vii) There is one group of four HII galaxies (the UM 462 group) without bright neighbours.

The results presented in the previous section and the conclusions above may lead to the following tentative interpretations:

• HII galaxies with companions tend to be of low luminosity and low velocity dispersions and have symmetric or regular morphology without any evidence of interaction. HII galaxies without companions have higher luminosity, higher velocity dispersion and tend to present distorted morphology. This is in contradiction with the expectation if the starburst in HII galaxies is triggered solely by tidal encounters.

• About 25 % of the sample are low luminosity, regular *and* isolated HII galaxies. For this subset there is no indication of the mechanism which could have triggered the present burst. In a hierarchical scenario of galaxy formation these gas rich dwarfs remain good candidates for being products of the surviving primordial low mass clouds in low density environments which failed to form luminous galaxies and are undergoing bursts of star formation at the present time (Silk 1983; Efstathiou & Silk 1983).

• W(H$\beta$) and oxygen abundance are the same for HII galaxies with or without companions. It is therefore unlikely that the environment influences the chemical evolution or the age of the starburst of HII galaxies. Total mass is more likely to be the relevant parameter, as observed for irregular galaxies (Skillman et al. 1989).

• We find puzzling the existence of a small group of HII galaxies (UM 462 group) with no bright galaxy in their immediate vicinity. This result, if real, raises an important additional question, i.e. how can star formation be synchronized in a few million years over distances of about 1-2 Mpc?

• The majority of HII galaxies are not formed as debris of interactions of giant galaxy pairs as suggested by Mirabel and collaborators.

• HII galaxies tend to populate very low density environments in the outskirts of clusters or loose group of galaxies.

The present results indicate that the general consensus that has been taken for granted in the last decade or more, namely that most starbursts are triggered by interactions with luminous companions, does not hold for most HII galaxies. Consequently, the findings of this study argue against models which try to relate the luminosity function of galaxies to the formation and evolution of structures in the universe assuming star formation to be controlled solely by dynamical processes (Lacey & Silk 1991, Lacey et al. 1993). The rate at which the gas is transformed into stars is more likely to be controlled by a combination of mechanisms: dynamical interactions, and/or internal processes such as the energy balance between the input by young stars and radiative cooling (White & Frenk 1991). The physical conditions which determine the mechanisms at play for different types of galaxies and different environments remain to be better understood.

A possible alternative comes from the work of Brinks (1990) who suggested that interactions with other dwarfs or intergalactic HI clouds could have triggered the current burst of star formation in HII galaxies. In a recent work Taylor, Brinks & Skillman (1993, 1994) using the VLA found at least one HI companion for 12 out of 19 HII galaxies. It remains to be seen if these low mass companions are capable of triggering a starburst and what is the relation if any between the presence of HI companions and HII galaxy properties. We believe that HI maps of the present sample of HII galaxies are essential to the improvement of our understanding of the triggering mechanisms in these small galaxies.

## ACKNOWLEDGMENTS

ET acknowledges his grant from CNPq/Brazil. We thank Harry Ferguson, Stacy McGaugh, Elena Terlevich and Elias